\documentclass[5p,authoryear]{elsarticle-var}
\usepackage{graphicx}
\usepackage{amssymb}
\usepackage{amsmath}
\usepackage{stfloats}
\usepackage[hyperindex,breaklinks]{hyperref}
\journal{New Astronomy Reviews}

\begin{document}

\begin{frontmatter}

\title{Modelling of mercury isotope separation\\
in CP stellar atmospheres: results and problems}


\author{A. Sapar}
\ead{sapar@aai.ee}
\author{A. Aret}
\author{L. Sapar}
\author{R. Poolam\"ae}

\address{Tartu Observatory, T\~oravere 61062, Estonia}

\begin{abstract}
Formation of anomalous isotope abundances in the atmospheres of chemically peculiar (CP) stars can
be explained by light-induced drift (LID). This effect is additional to the radiative acceleration
and appears due to systematic asymmetry of radiative flux in partly overlapping isotopic spectral line
profiles. LID causes levitation of an isotope with a red-shifted spectral line and sinking of an isotope with
a blue-shifted line, generating thus diffusive separation of isotopes.
We have studied diffusion of mercury as a typical well-studied isotope-rich heavy metal.
Our model computations show that in mercury-rich quiescent atmospheres of CP stars
 LID causes levitation of the heavier mercury isotopes and sinking of the lighter ones.
Precise quantitative modelling  of the process of isotope separation
demands very high-resolution computations and the high-precision input data,
 including data on hyperfine and isotopic splitting
 of spectral lines, adequate line profiles and impact cross-sections.
Presence of microturbulence and
  weak stellar winds can essentially reduce the effect of radiative-driven diffusion.

\end{abstract}


\end{frontmatter}

\section{Introduction}
\label{intro}
The problem of specifying the physical processes which cause the observed
drastic overabundance of several heavy metals and anomalous abundance ratios
of  their isotopes in the atmospheres of chemically peculiar (CP) stars,
especially of chemically peculiar mercury--manganese (HgMn) stars,
has been on agenda already for several decades.
At the present time almost nobody doubts,  that atomic diffusion is responsible for a large
part of the abundance variations observed in CP stars.
About 40 years ago Michaud demonstrated in his
pioneering papers \citep{1970ApJ...160..641M, 1976ApJ...210..447M}
that for  spectral line-rich ions of trace elements
the gradient of the radiative
pressure, i.e. the upwards directed  radiative acceleration,
 essentially exceeds
the downwards directed gravity. This circumstance produces the phenomenon of
accumulation of spectral line-rich heavy metals in the stellar atmosphere.
Further numerous papers have confirmed the result.

Stellar evolution models with radiative-driven diffusion have been constructed by Michaud and his colleagues
for AmFm stars \citep{2000ApJ...529..338R, 2005ApJ...623..442M},
Population II stars \citep{2002ApJ...568..979R, 2002ApJ...580.1100R, 2002ApJ...571..487V}
and horizontal branch (HB) stars \citep{2007ApJ...670.1178M, 2008ApJ...675.1223M}.
HB stars are very closely related to HgMn stars.
These models do not include detailed atmospheric modelling and assume chemically homogeneous mixed outer region.
Radiative accelerations have been found using so-called ``diffusion approximation''
for photon flux  \citep{1927MNRAS..87..697M}. This approximation is valid in optically thick medium of
stellar interiors where $\tau \gg 1$.
Evolutionary computations show \citep[see][Fig.~1]{2008CoSka..38..103M} the high trend of
 cumulation of heavy metals in the outer layers of stars. Line-rich metals are
swept up throughout the stellar interior by upwards directed radiative push.
Radiative acceleration on the extremely line-rich low ionization stage
ions of heavy metals can exceed downwards directed gravity up to 3 dex.

Stellar evolution models have been relatively successful at explaining peculiarities of AmFm stars.
However, not all abundance anomalies of HgMn and HB stars can be reproduced.
Isotopic anomalies observed in HgMn stars also remained unexplained.
\citet{2008CoSka..38..103M} suggested that additional separation occurs in the atmospheric regions.

Observational evidences of abnormal isotope ratios in CP stellar atmospheres
have been cumulated during last decade owing to
 the high-resolution and high signal-to-noise exposures of the HST and
ground-based telescopes with similar capabilities.
Although the overall picture of isotope variations is complex \citep[see for review][]{2008CoSka..38..291C}, there is a general regularity for heavy elements
to show overabundance of heavier isotopes and for light elements -- overabundance of lighter isotopes. For example, isotopic abundances of mercury range
from the terrestrial mixture to the virtually pure $^{204}$Hg while overall overabundance of mercury is 4--5 dex \citep{2003A&A...402..299D}.
In spite of numerous attempts, strong isotopic anomalies
observed in the atmospheres of chemically peculiar stars
have not found acceptable explanation yet. Separation of isotopes requires some additional physical mechanism,
because the radiative push is almost the same for all isotopes of the particular element.

A possibility for breakthrough was elucidated by
\citet{AtutovShalagin1988}, who, based on the results of laboratory laser experiments,
proposed a new physical mechanism, called light-induced drift (LID), for explaining
the isotope separation in the quiescent atmospheres of CP stars.

\section{Concept of LID and its main formulae}
\label{LID}
The proposal by \citet{AtutovShalagin1988}  inspired us
\citep{2002AN....323...21A,2008CoSka..38..273S,2008CoSka..38..445S} to investigate the possibilities to unveil
the enigmatic appearance of isotope anomalies in the CP stellar atmospheres using LID phenomenon.
This new mechanism is fully based on the asymmetry of spectral lines.
It works effectively if spectral line profiles of a studied element are systematically
asymmetrical. This is the case for partly overlapping mutually shifted isotopic lines.

The  absorption of the radiative flux,
which is asymmetrically distributed in the blue and red wings of a spectral line, generates
anisotropy of excitation rate of the atomic particles participating in thermal motion.
For example, larger radiative flux in the red wing of a spectral line
gives more excited particles among those moving downward in the atmosphere than among those moving upward.
Particles in excited states have larger collision cross-sections than particles in the ground
or lower excitation states and thus the mobility of particles in the higher excited states is smaller.
If particles survive in the excited states until the next collision, then their
free paths are shorter than free paths of the unexcited particles. This
means that radiative flux asymmetry in the spectral lines triggers a diffusion phenomenon.

The first task was to find
the formulae describing this phenomenon. We succeeded to reduce the LID to the
equivalent acceleration to be added to the usual radiative acceleration
\citep{2002AN....323...21A, 2008CoSka..38..445S}.

As well known, the usual radiative acceleration due to electron transition
from lower level $l$ to upper level $u$ of ion $j$ can be found as
\begin{equation}
a^\mathrm{rad}_j = \frac{\pi}{m_jc}\int^\infty_0 \! X_{j,l} \sigma_{ul}^0 V(u_\nu, a) \mathcal{F}_\nu d\nu~.
\label{arad}
\end{equation}
Here $m_j$ is the mass of the light-absorbing ion, $X_{j,l}$ is
state $l$ population fraction, $\pi \mathcal{F}_\nu$ is total
monochromatic flux and $\sigma_{ul}^0$ is the absorption cross-section in transition \( l\rightarrow u \):
\begin{equation}
\sigma_{ul}^0 = \frac{\pi\,e^2 f_{ul}}{m_e c \Delta \nu _{D}}~,
\end{equation}
where $f_{ul}$ is oscillator strength and $\Delta \nu _{D}$ is Doppler width of the
spectral line.
Normalized frequency distribution in a spectral line is the Voigt function,
 being the convolution of the Lorentz and Doppler profiles, i.e.
\begin{equation}
      V(u_\nu,a) = \frac{a}{\pi^{3/2}}\int\limits_{-\infty}^\infty \frac{e^{-y^2}}{(u_\nu - y)^2 + a^2 } dy~.
\label{voigtfunc}
\end{equation}
The argument of the Voigt function is the dimensionless frequency
 $u_{\nu }=\Delta \nu/\Delta \nu _{D}$,
 its parameter $a=\Gamma_{ul}/(4\pi \Delta \nu _{D})$ is the ratio of characteristic widths of Lorentz
  and Doppler profiles.
Integration is carried out over  the dimensionless velocity
$y=v/v_{T}$, where the thermal velocity $v_{T}=\sqrt{2kT/m_j}$.

 LID appears  due
to difference of collision  cross-sections in different quantum
states. LID efficiency depends on the difference between collision rates in upper and lower state
and also on the probability of particle to stay in the excited state until the next collision.

Based on these  conceptual standpoints we have derived
 the following
expression for the equivalent acceleration due to LID in the transition $ l\rightarrow u$
 of ion~$j$:
\begin{equation}
a_j^\mathrm{LID}=\varepsilon q\frac{\pi}{m_jc}\int\limits_0^\infty X_{j,l}\, \sigma^{0}_{ul}\,
\frac{\partial V(u_\nu, a)}{\partial u_\nu}\mathcal{F}_\nu d\nu~.
\label{alid}
\end{equation}
Coefficient $q$ is the ratio of the mean momentum of the atomic particles and the photon momentum:
\begin{equation}
q=\frac{m_jv_Tc}{2 h\nu}=\frac{m_jv_T}{2}:\frac{h\nu}{c} ~.
\end{equation}
Due to the large value of $q \approx 10,000$, LID can be important even in
the case of moderate asymmetry of radiative flux in the spectral
line profiles. Under favourable conditions LID may be several orders of magnitude larger than
usual radiative acceleration in line.
Efficiency of LID $\varepsilon$ is given by
\begin{equation}
 \varepsilon=\frac{C_u-C_l}{C_u} \cdot \frac{C_u}{A_u+C_u} = \frac{C_u-C_l}{A_u+C_u}~,
 \label{lideff}
\end{equation}
where $A_u$ is the total rate of radiative transitions from the upper state.
The first term in Eq.(\ref{lideff}) is the relative difference between collision rates in upper and lower states.
The second term expresses the probability that
 during the free flight after photon absorption de-excitation does not happen.
Unlike the amplification coefficient $q$, the value of $\varepsilon$
can be estimated only with rather low precision. Characteristic
values of $\varepsilon$ are from about $10^{-3}$ in the outer atmospheric
layers of the CP stars to about $1/4$ in the deep layers.

Quantum state life-times have
been estimated similarly to the Kurucz spectrum synthesis program SYNTHE \citep{1993KurCD..18.....K}
according to the formulae given in the book by \citet{1974bmae.book.....K}. Radiative and Stark damping
constants by \citet{1999ApJ...512..942P} were used for the lines given in their paper.
Cross-sections of mercury
collisions with buffer gas (H + He) particles in the upper state $u$ are larger than in the lower state $l$.
These collisions are predominantely elastic. The collision cross-sections were calculated adopting
the quasi-hydrogenic approximation. For ion--ion impacts also the additional Coulomb repulsive cross-sections were
taken into account.  Diffusion coefficients were calculated according to the  \citet{1995A&A...297..223G}.
 These approximations allow to study general trends of isotope separation
due to LID. More detailed computations of cross-sections and quantum
state life-times are necessary in the future studies, but general picture
should remain the same.

The total effective acceleration, acting on atomic particle due to both the
radiative force and the LID is given by
$a^\mathrm{t}_j = a^\mathrm{rad}_j +a^\mathrm{LID}_j$.
Thus, the effect of the light-induced drift can be incorporated by replacing the
Voigt function $V(u_{\nu },a)$ in the expression of radiative acceleration
 Eq.(\ref{arad}) by
\begin{equation}
 W(u_{\nu},a)=V(u_{\nu},a)+ \varepsilon q \frac{\partial V(u_{\nu},a)}{\partial u_{\nu}}~.
\end{equation}
Once again, we would like to emphasize, that the LID can be treated as acceleration generated by
additional specific force.

Light-induced drift is most effective in the intermediate layers of stellar atmospheres.
In high layers of atmosphere or above it
LID efficiency is small because of long free paths of the particles, i.e. probability of particle to return
to the lower state before the next collision is larger.
In stellar interiors LID is inessential compared to the usual radiative acceleration due to other effects:
first, isotopic splitting of lines of high ions (e.g. Hg~III, IV etc.) is so large that isotopic lines do not overlap and
consequently do not give a systematic asymmetry of the lines generating large LID;
second, radiative flux becomes local and asymmetry of flux diminishes.

\section{On the isotopic and hyperfine splitting of spectral lines }
\label{splitting}
The efficiency of the LID phenomenon is governed by the finest details of spectral line
splitting, including their hyperfine and isotopic structure, which determines the
regularities in the spectral line overlapping and thus the degree of their asymmetry.
General expression for  energy shift of quantum states of isotope with mass $M$ is
sum of the normal (virial), specific (the momentum correlation of nucleons)
and field (nuclear volume) shifts. Symbolically, it can be expressed in the form
\begin{equation}
\Delta E_M=-\frac{m_e}{M}E_0+E_s+E_A.
\end{equation}
The first term here is the   normal shift, being, as seen from the expression,
large for light element isotopes and small for the heavy ones.
The specific shift gives essential contribution only for the medium weights
 atoms  with $Z=10, \dots, 20$, and its computation is rather complicated. The third term is the nuclear volume
 shift due to symmetric charge distribution in the nucleus,
 depending on the integrated product of the mean charge density in nucleus
  and on the square radius of nucleus, i.e.  approximately holds
  \begin{displaymath}
   E_A\propto\langle\rho_e r_M^2\rangle~,
   \end{displaymath}
    where $\rho_e$ is charge density in the nucleus and $r_M$
   is current value of the radius.
   As well known, only the
  s-state electrons have non-zero charge in the centre of nucleus. Therefore, the volume shift has
  largest contribution for s-electrons and it is essentially lower for the other states.

 If nucleus  is a sphere
with radius   $R$ and charge distribution is uniform  in it, then
\begin{displaymath}
R^2=\frac{5}{3}\langle r^2_M\rangle~. 
\end{displaymath}
For atoms with mass number $A>16$ holds  approximately (in fm units)
$
R =1.1154 A^{1/3}
$
and thus
\begin{equation}
E_A \propto\rho_e A^{2/3}.
\label{volshift}
\end{equation}
Naturally, this  formula gives only a rough approximation for the  volume shifts.
As seen from the Eq.(\ref{volshift}), the volume shift
grows  monotonously  with atomic weight contrary to the normal shift, which diminishes for
the heavier nuclei.

 In the isotopes with odd number of nucleons presence of  magnetic dipole momentum of  nucleus and
anisotropy of the charge distribution
generate  the atomic energy level  shifts due to total
 magnetic dipole and electric quadrupole momenta of  nucleus, correspondingly.
 The spectral line shifts of the bound states which
 appear  due to magnetic dipole, are specified by
\begin{equation}
W_{H2}=\frac{A}{2}C
\end{equation}
and due to quadrupole momentum
\begin{equation}
W_{E4}=B\frac{\frac{3}{4}C(C+1)-I(I+1)-J(J+1)}{2I(2I-1)2J(2J-1)},
\end{equation}
where
 \begin{equation}
C=F(F+1)-I(I+1)-J(J+1).
\end{equation}
In these formulae $I$ and $J$  are the nuclear and total electron angular momentum quantum numbers,
respectively, and $F$ is their vector sum.
The quantities $A$ and $B$ depend correspondingly on the  magnetic dipole
and electric quadrupole momenta of the atomic nucleus.

\section{Separation of mercury isotopes in CP stellar atmospheres}

\begin{figure*}[b]
  \centering
  \includegraphics[viewport=0 15 330 240, clip, width=.30\textwidth]{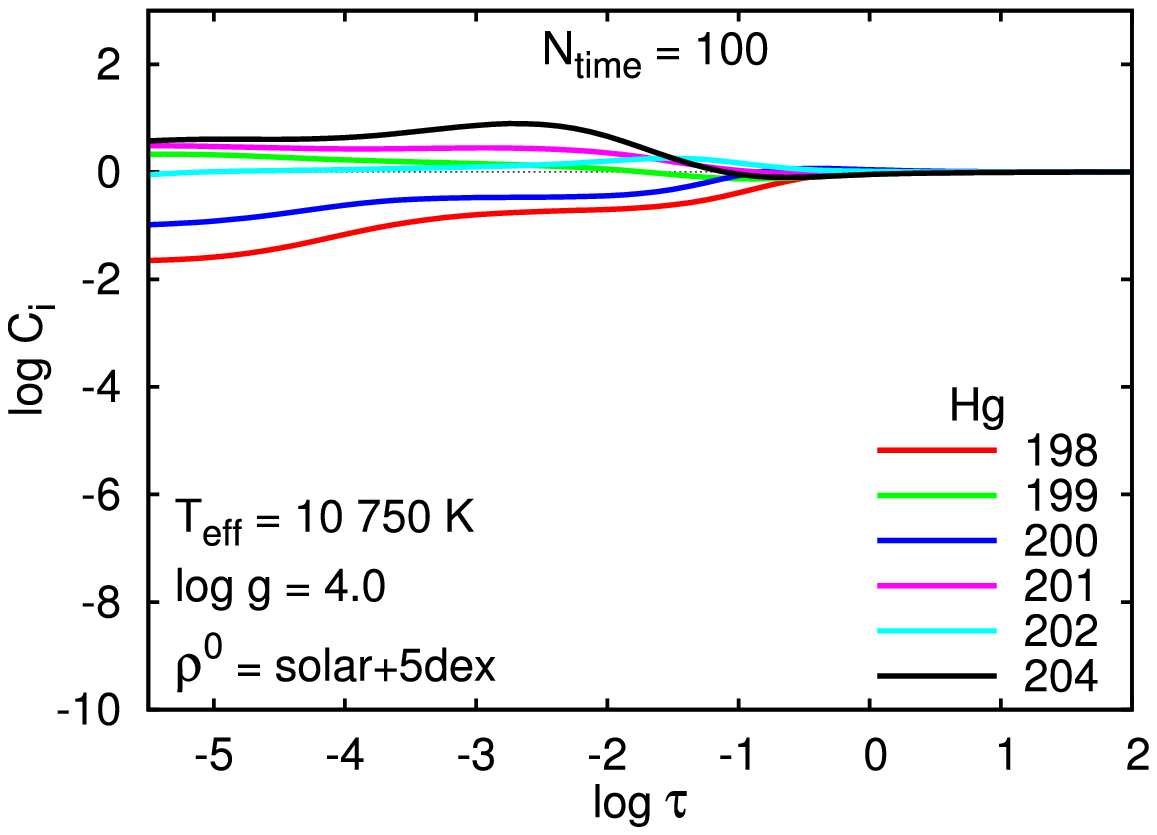} \hspace{10pt}
  \includegraphics[viewport=0 15 330 240, clip, width=.30\textwidth]{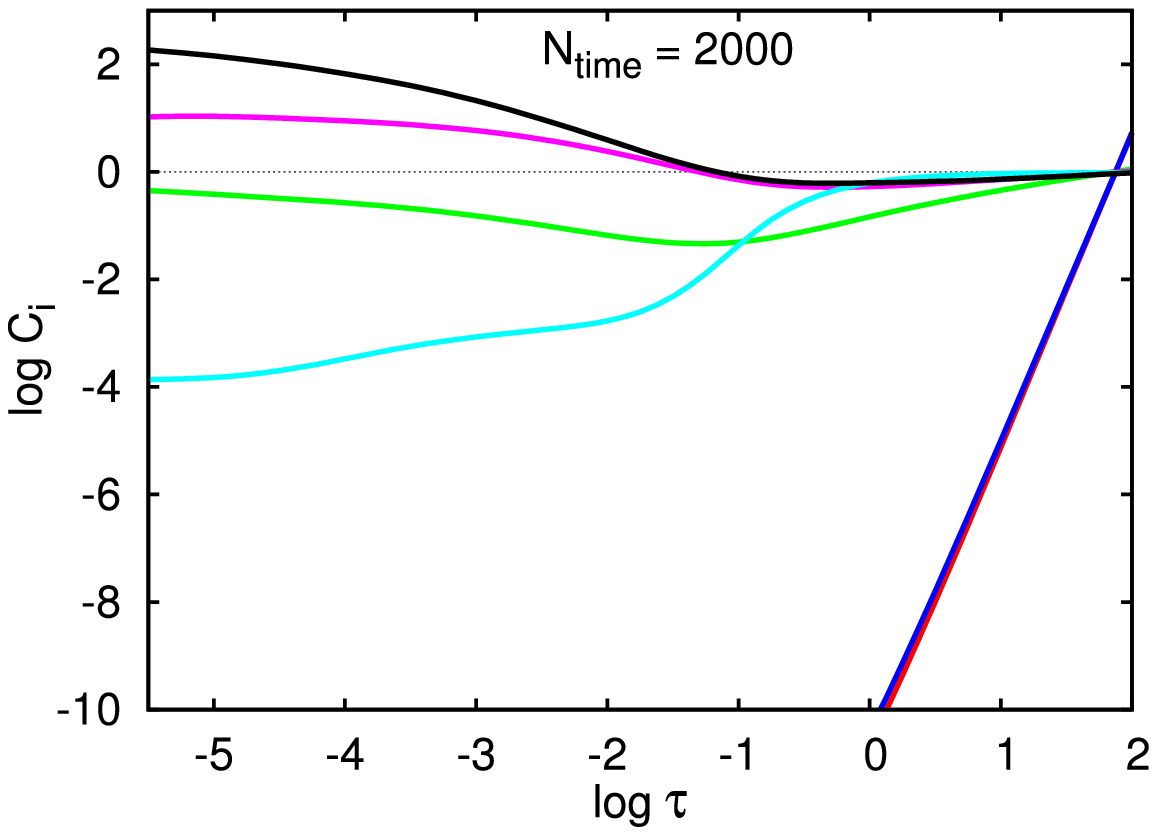} \\[0pt]
  \includegraphics[viewport=0 15 330 240, clip, width=.30\textwidth]{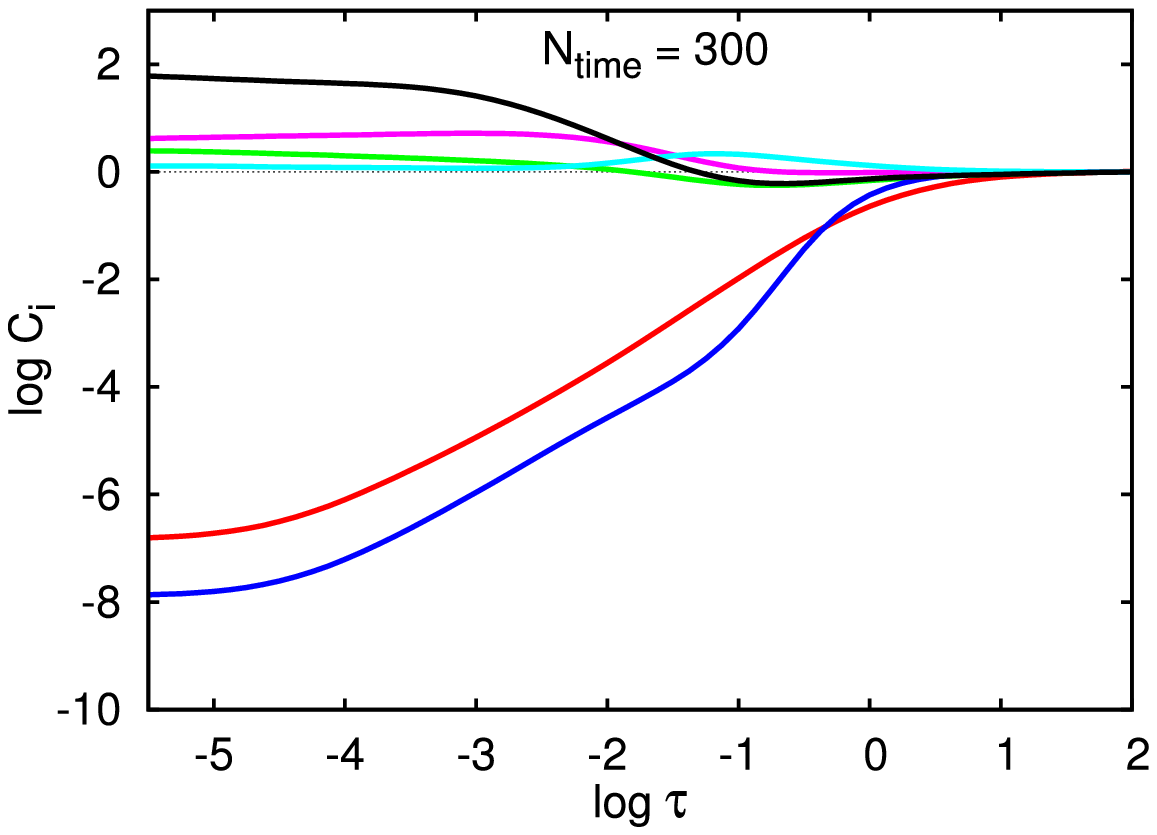} \hspace{10pt}
  \includegraphics[viewport=0 15 330 240, clip, width=.30\textwidth]{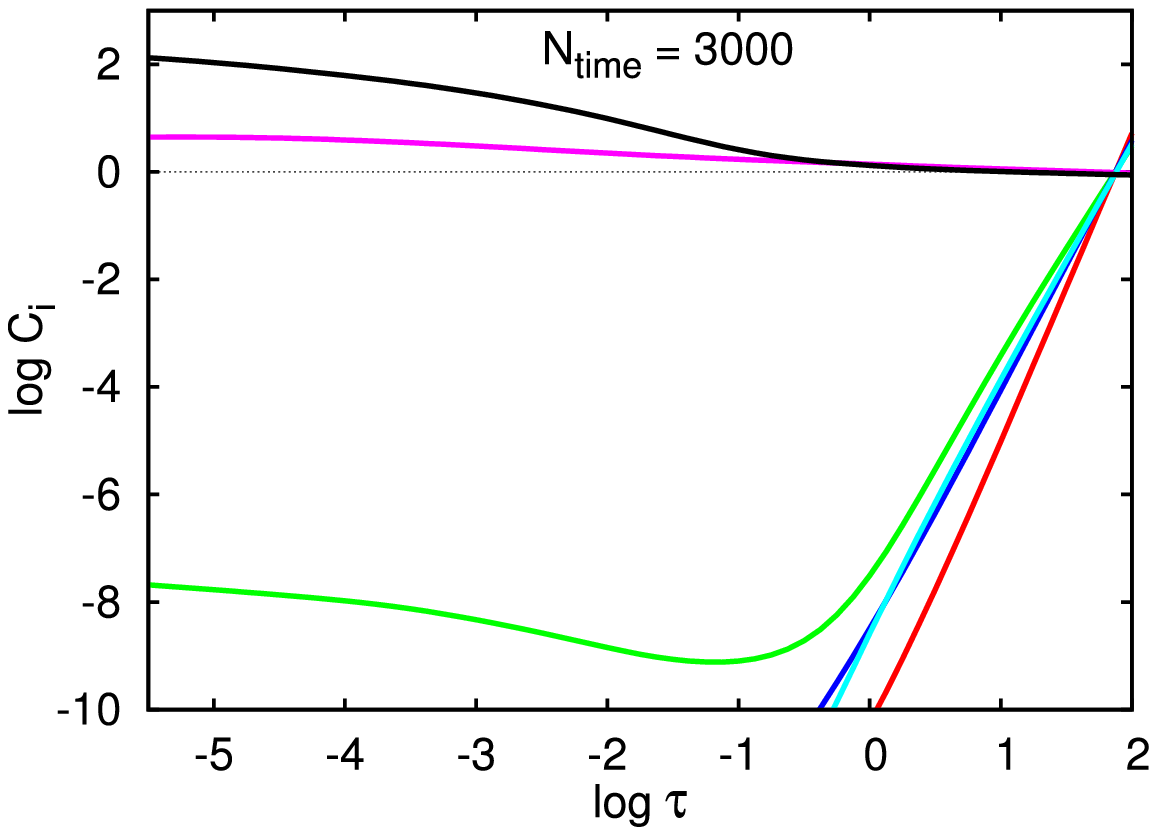}\\[0pt]
  \includegraphics[viewport=0 15 330 240, clip, width=.30\textwidth]{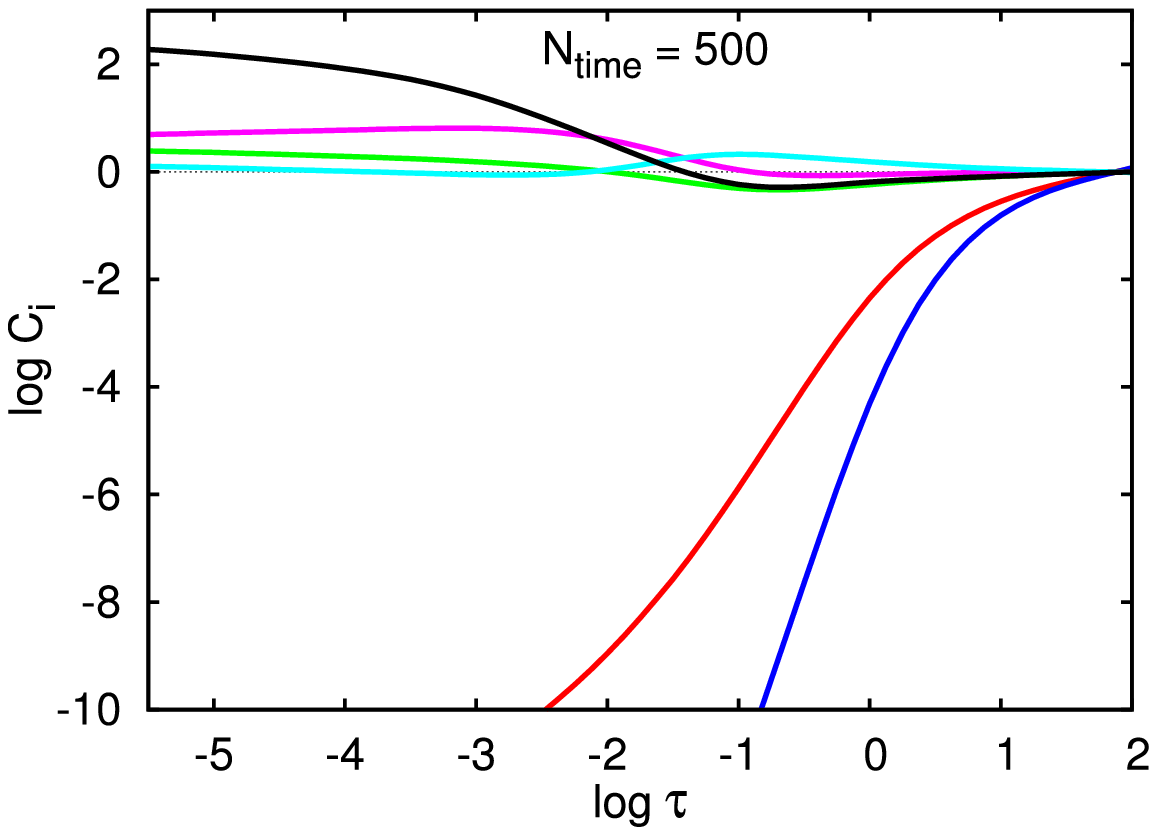} \hspace{10pt}
  \includegraphics[viewport=0 15 330 240, clip, width=.30\textwidth]{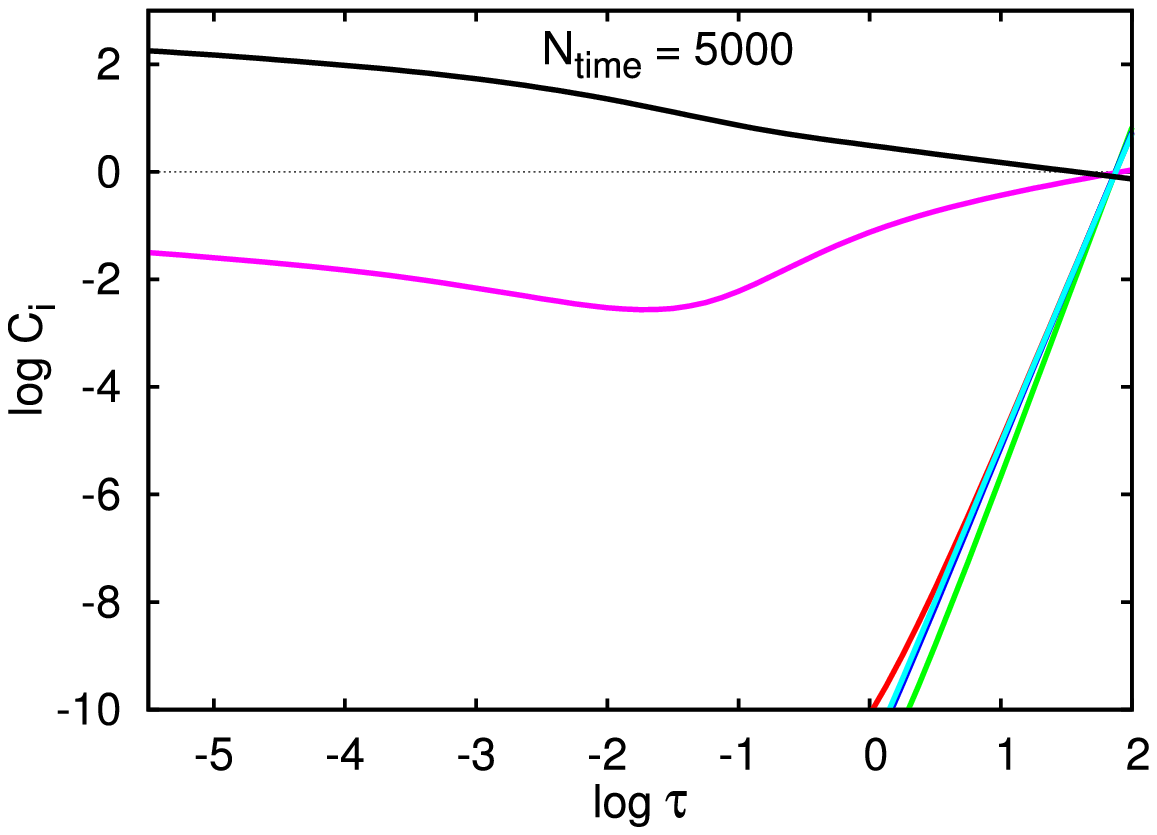}\\[0pt]
  \includegraphics[viewport=0 0 330 240, clip, width=.30\textwidth]{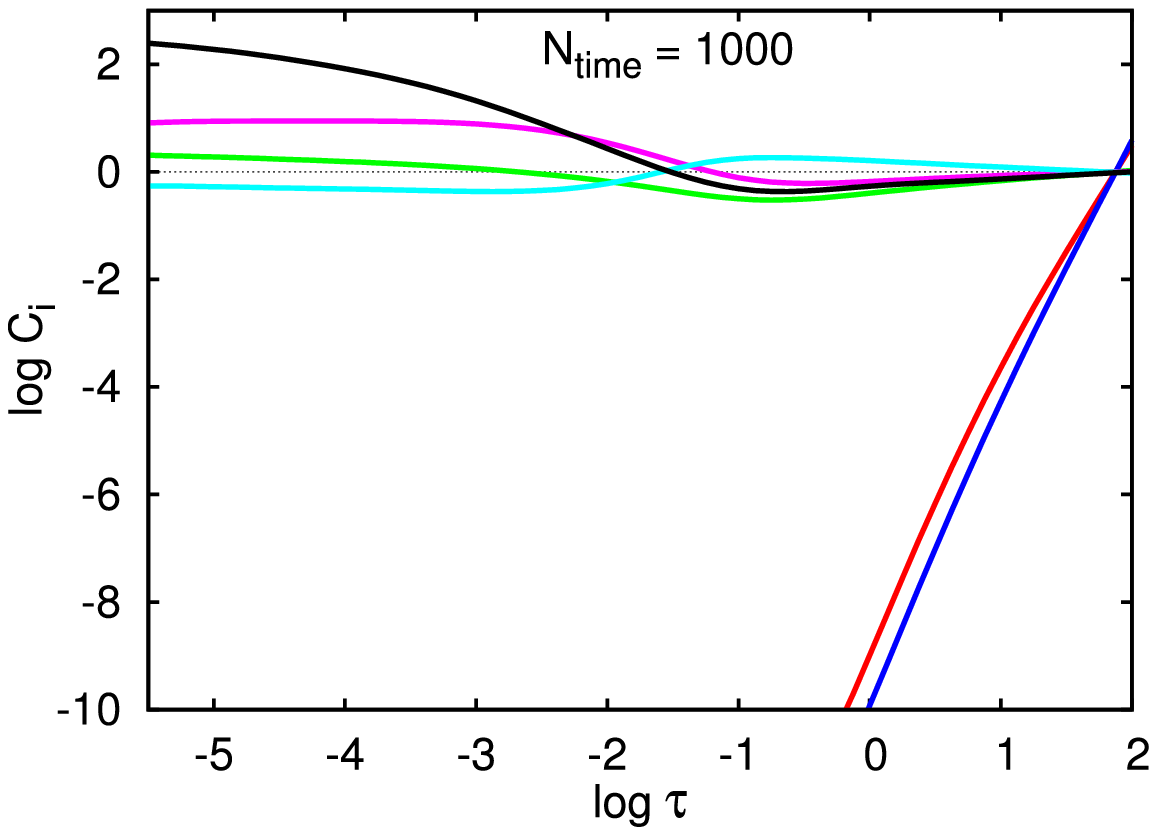} \hspace{10pt}
  \includegraphics[viewport=0 0 330 240, clip, width=.30\textwidth]{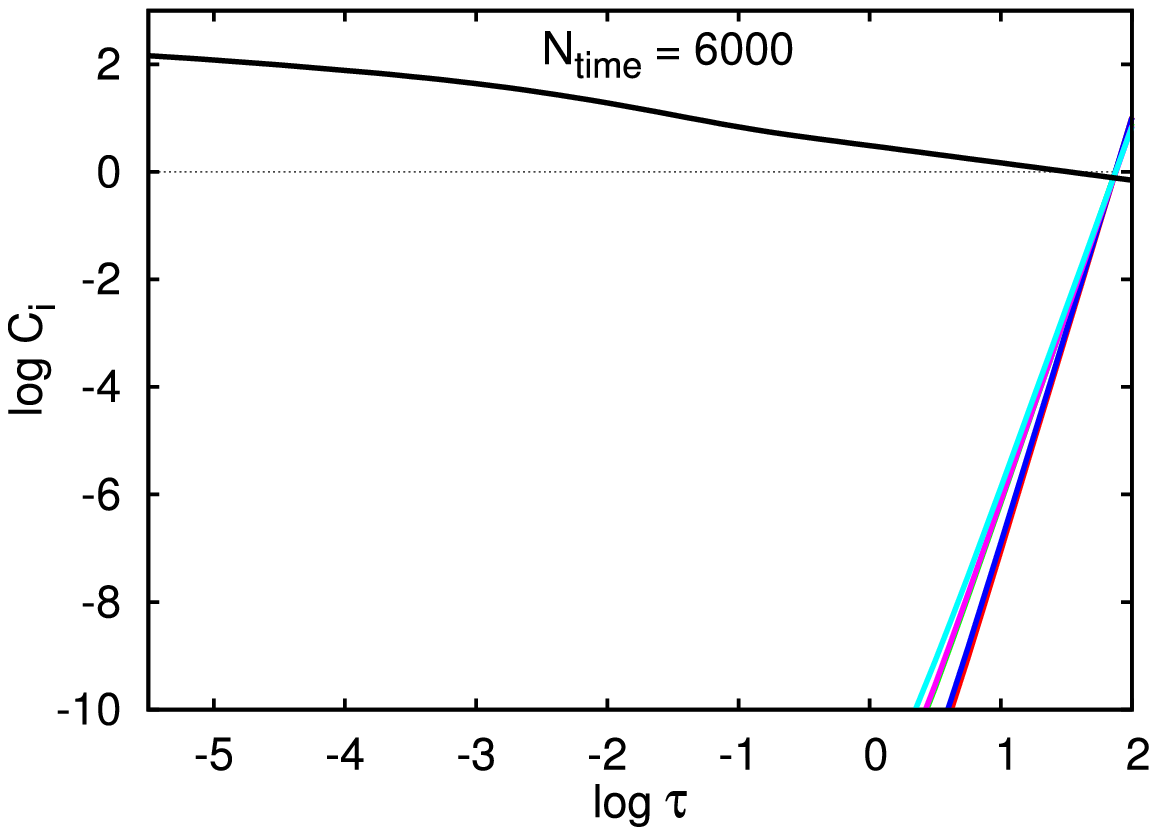}
  \caption{Evolutionary changes of mercury isotope concentrations relative to the initial ones $C_i = \rho_i/\rho_i^0$ in model atmosphere with
$T_\mathrm{eff}= 10~750~\hbox{K}$, $\log g=4$, $V_\mathrm{rot}=0$, $V_\mathrm{turb}=0$ and
initial Hg abundance solar~+~5~dex with solar system isotope ratios.}
   \label{colog10750}
\end{figure*}

   Generation of elemental and isotopic peculiarities in the atmospheres of
   CP stars is described by the following scenario.
   During stellar evolution line-rich heavy metals are pushed to the atmosphere from stellar interiors
    due to the large expelling radiative force which can
    exceed the gravity by several dex \citep{2008CoSka..38..103M}.
    However, this radiative drive leaves solar isotope ratios unaltered.

   Isotope separation takes place in stellar atmospheres, where
   the local diffusive radiative transfer is
    replaced by non-local radiative transfer. This favours generation of
    asymmetrical radiative flux in profiles of spectral lines.
    Flux asymmetry within spectral line profiles of atoms and ions
    generates light-induced drift, which is not random but
    cumulative for isotopes, because
    isotopic splitting of spectral lines creates systematic asymmetry in overlapping line profiles.

Study of evolutionary separation of heavy metals, predominantly of Hg,
and their isotopes due to LID in the atmospheres of CP stars
has been an essential topic of our investigations during about last decade.
Computer code SMART \citep{2003ASPC..288...95S,2007smma.conf..236S,2008IAUS..252...41A},
composed primarily for modelling of stellar atmospheres and stellar spectra,
has been supplemented with additional software blocks
 for computation of evolutionary scenarios of
diffusive separation of isotopes of chemical elements due to radiative acceleration, LID and gravity.
Mercury has been chosen for the modelling mainly because of two reasons. First, it
is a typical isotope-rich heavy metal. Second, mercury is highly overabundant in HgMn stars
and its isotopic mixture
has been determined from observations for more than 30 stars \citep{1999ApJ...521..414W, 2003A&A...402..299D}.

It has turned out that formation of
anomalous isotope abundances in stellar atmospheres, including the
dominance of the heaviest isotope,
cannot be explained by the diffusion theory without including LID mechanism.

Since LID is sensitive to the shape of spectral lines, synthetic spectra must be computed with high
resolution at all layers of model atmosphere.
Radiative flux in optically thin medium of stellar atmosphere has to be found
resolving the radiative transfer equation in detail.
This is more difficult and time-consuming than computing the flux
according to the so-called diffusion approximation, valid
in optically thick media  of stellar interiors.
Precise data on hyperfine and isotopic splitting
 of spectral lines and cross-sections of various physical processes are also necessary
in LID computations.

Spectral line data for mercury have been compiled  using different sources and improved by
adding isotopic splitting to all available Hg lines. Most of the line data have been taken
from the lists by \citet{1995KurCD..23.....K} and Vienna atomic line database
 VALD \citep{2000BaltA...9..590K}. Hyperfine and isotopic structure of Hg lines and their oscillator strengths
available in papers by \citet{1999ApJ...512..942P} and \citet{1997A&A...319..928S}
were used to improve our line list. For other lines isotopic splitting was calculated using relative shifts
found by \citet{StriganovDontsov1955}, scaled to units $[202-200]\,$=$\, 1$,
giving for other splitted line components
$[198-200]\,$=$\, -0.94$, $[199-200]\,$=$\, -0.80$, $[201-200]\,$=$\, 0.30$,
 $[204-200]\,$=$\, 1.98$.
 For scaling these relative shifts to wavenumbers,
  we have taken into account that $[202-200]\,$=$\, 0.179$~cm$^{-1}$ for Hg~I,
  $[202-200]\,$=$\, 0.508$~cm$^{-1}$ for Hg~II and $[202-200]\,$=$\, 0.600$~cm$^{-1}$ for Hg~III.
Compiled line list contains about 700 resonance and low excitation spectral lines in the wavelength range approximately 800--12,000~{\AA}
  for Hg~I,  Hg~II and Hg~III, i.~e. for ion species, which are  most important for LID.

In this paper we present some results of modelling the evolutionary separation of mercury isotopes
 and discuss some problems on the way of further studies.
 The problems of LID modelling in the quiescent atmospheres of slowly rotating CP stars
 involve the computation of line strengths, formation of line profiles
  due to different microphysical interaction processes, but also formation of microturbulence and
  weak stellar winds as the phenomena, reducing the diffusion. In addition, presence of entangled magnetic field
  can affect the formation of anomalous isotope abundances, mainly due to
  Zeeman splitting of spectral lines.

\begin{figure}
  \centering
  \includegraphics[viewport=0 0 330 250, clip, width=.30\textwidth]{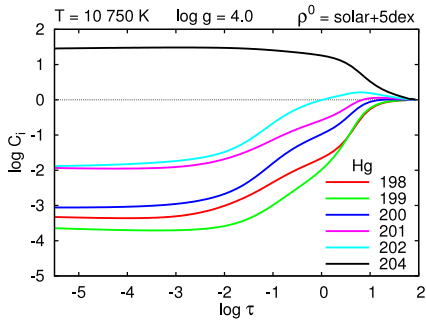}
  \caption{Equilibrium mercury concentrations $C_i = \rho_i/\rho_i^0$ in model atmosphere with microturbulence.
Microturbulent diffusion coefficient is assumed to be 50 times the atomic diffusion coefficient.
Model parameters: $T_\mathrm{eff}= 10~750$~K, $\log g=4$, $\rho_i^0=$~solar~+~5~dex.
}
   \label{equilibr}
\end{figure}

Example of a computed evolutionary scenario of separation of
mercury isotopes due to LID in mercury-rich (solar + 5 dex) main sequence stellar atmosphere
($T_\mathrm{eff}= 10~750~\hbox{K}$, $\log g=4$, $V_\mathrm{rot}=0$, $V_\mathrm{turb}=0$)
is given in Fig.~\ref{colog10750}.  Macromotions that can suppress diffusion effects have been ignored.
Consequent frames show relative changes of mercury isotopic abundances $C_i = \rho_i/\rho_i^0$ with time
throughout the atmosphere ($-6 < \log\tau < 2$).
Lighter even-A isotopes sink first. Odd-A isotopes are supported longer in the atmosphere
due to hyperfine splitting of their spectral lines.
As stated above, the hyperfine splitting of atomic states is essential only for odd-A isotopes.
Hyperfine splitting mixes the order of isotope lines and makes
the picture of evolutionary isotope separation more complicated.
However, as time passes, all lighter isotopes sink and
 only the heaviest isotope $^{204}$Hg remains in the atmosphere.
Separation in upper layers of the atmosphere proceeds essentially faster than in
deep layers. We would like to emphasize that present evolutionary scenario has been computed
for perfectly quiet stellar plasma, without microturbulence, stellar wind or any other mixing process.
Presence of macromotions in the atmosphere drastically slows down the diffusion and also reduces resulting
abundance gradients.

Final equilibrium isotope separation profiles throughout the atmosphere in the presence of microturbulence (Fig.~\ref{equilibr})
 have  been iteratively computed for the same model atmosphere as the evolutionary scenario presented in Fig.~\ref{colog10750}.
Microturbulent diffusion coefficient has been assumed to exceed the atomic diffusion coefficient 50 times.
The heaviest  isotope $^{204}$Hg is also strongly dominant,
however abundance gradients are not as steep as in Fig.~\ref{colog10750}.

\begin{figure}[b]
  \centering
  \includegraphics[width=.30\textwidth]{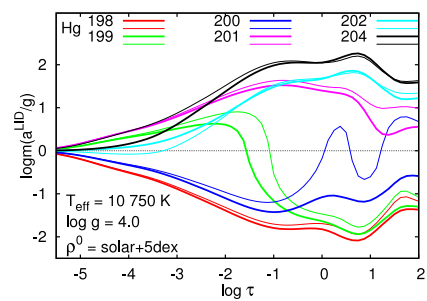}
  \caption{Effect of line blendings on LID of Hg isotopes.  Accelerations
  of the Hg isotopes due to LID at the first time step (i.e. with homogeneous Hg abundance) are shown in the modified logarithmic scale
  $\mathrm{logm}\left(\dfrac{a}{g}\right)   = \mathrm{sign}(a) \log\left(\left|\dfrac{a}{g}\right| + 1\right)$.
  Thick lines: only Hg isotopic lines, no blendings with lines of other elements;
  thin lines: lines of all elements are present.}
   \label{Hg-vs-all}
\end{figure}

Light-induced drift is generated when the radiative flux is asymmetric within the line widths of
the light-absorbing ion. This asymmetry mainly appears
due to line blendings, which are usually random. Thus, a direction and value of the LID in each spectral line  is also random and total LID, obtained
summing over all spectral lines of studied element, is minor. Total value of LID can be large only if there is a systematic asymmetry of spectral lines,
what is the case for the overlapping isotopic lines. Thereby LID of mercury isotopes is governed by mutual overlaps of their spectral lines,
while blendings with lines of other elements are less important, because they statistically cancel out. Calculations of LID due to overlapping lines of mercury isotopes in the presence and without lines of
other elements confirm this conclusion (Fig.~\ref{Hg-vs-all}). Blendings with lines of other elements certainly affect values of $a^\mathrm{LID}$, but do not change the main regularities defined by mutual overlaps of Hg isotopic lines.

\section{Hydrogenic line profiles}

Although details of mercury lines are crucial for
computation of LID acceleration, also profiles of all strong lines,
which overlap with mercury lines, should be computed with high precision to obtain more realistic picture of isotope separation.
This concerns also the hydrogen spectral line series.
 As we have shown \citep{2006BaltA..15..435S},
 the convolution of Holtsmark, Lorentz and Doppler
profiles in the result of two analytical integrations reduces to the sum of three contribution terms, namely
\begin{equation}
\Phi(\beta)=\chi(\beta)+\Lambda(\beta )+\Delta(\beta ),~~~~~~~~~
\end{equation}
where
\begin{equation}
\chi(\beta)=\frac{2}{\pi}\int_0^\infty\beta x\sin(\beta x)\varepsilon(x)dx,
\end{equation}
\begin{equation}
\Lambda(\beta )=\frac{2L}{\pi}\int_0^\infty x \cos(\beta
x)\varepsilon(x)dx,
\end{equation}
\begin{equation}
\Delta(\beta)=\frac{4D^2}{\pi}\int_0^\infty
x^2\cos(\beta x)\varepsilon(x)dx
\end{equation}
and
\begin{equation}
\varepsilon(x)=\exp\left(-x^{3/2}-Lx-D^2x^{2}\right).
\end{equation}
In these formulae $\beta=(\nu-\nu_0)/\Delta\nu_S$, $L={\Delta\nu_L/\Delta\nu_S}$ and $D={\Delta\nu_D/ 2\Delta\nu_S}$.
Here $\Delta\nu_D$, $\Delta\nu_L$ and $\Delta\nu_S$ are the
widths parameters of Doppler, Lorentz and Holtsmark line profiles, respectively.

Another possibility is numerical integration of
    the convolution of  analytical high-precision approximation expressions
  of Holtsmark  and Voigt functions \citep{2006BaltA..15..435S}.
    The asymptotic series expansions
    for the small and large values of argument $\beta$
have been found by us  as a contribution to this approach.
For the intermediate region the exponential term can be expanded into series
followed by integration over circle.

\begin{figure}
  \centering
  \includegraphics[viewport=0 0 326 225, clip, width=.30\textwidth]{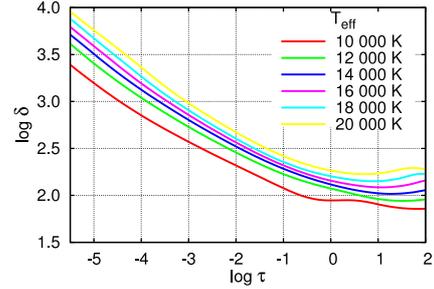}
  \caption{Number of charged particles in the Debye sphere.
  Large Ecker coefficient values $\delta > 50$ show that
 Holtsmark distribution is valid throughout stellar atmospheres.}
   \label{debye}
\end{figure}

    For main sequence stars Holtsmark distribution shielded by the Debye radius, i.e. the Ecker
   correction is not essential, because  the value of screening parameter, approximately estimated by
\begin{equation}
    \delta=\frac{4\pi R_D^3}{3} N,
\end{equation}
    where $R_D$ is the Debye radius
\begin{equation}
    R_D=\left(\frac{kT}{4\pi e^2N}\right)^{1/2}
\end{equation}
    and
     $N$ is the number density of charged particles, has large values up to deep atmospheric layers.
     This is illustrated in Fig.~\ref{debye}. As known, Holtsmark distribution holds well if $\delta$, being the number
     of particles inside the Debye sphere, exceeds about 50.

      As mentioned earlier, an important problem for modelling of stellar atmospheres
       is to find high-precision profile function for hydrogen lines.
The Model Microfield Method (MMM) initially developed by
\citet{1971JQSRT..11.1753F} and \citet{1971JQSRT..11.1767B} and further
elaborated by Stehl\'{e} et al.
\citep[e.g.][]{1990JQSRT..44..135S,1993A&A...271..348S} takes
statistically into account also the ion dynamics effects. Tables for
the Lyman and Balmer \citep{1994A&AS..104..509S,1999A&AS..140...93S},
Paschen \citep{1996PhST...65..183S} and Brackett lines
\citep{2009arXiv0907.0635S} have been calculated and are available
for astrophysical applications. The main difference between MMM and
Holtsmark profiles lies in the line centre. The profile function
obtained with MMM has a non-zero plateau at small argument values
while the Holtsmark distribution approaches zero. However, the total
contribution of blendings with hydrogen lines to the LID of mercury
isotopes turned out to be minor and eventually also the MMM
correction does not essentially affect our results.

 Splitting of H spectral lines due to the linear Stark effect, where the number of splitted line components
 grows rapidly towards the higher members of the series,
 also must be further taken into account.
   This problem is connected with  determination of  dissolution rates
  of high Rydberg lines. The state survival fractions have been specified at least
   in three different ways, giving
  the expressions in the form
  \begin{equation}
  W_n=CN^kn^m~,
\end{equation}
    where $N$ is the number density of charged particles
    and $n$ is the current value of the principal quantum  number. These methods are:
    \begin{itemize}
\item[--]  Inglis--Teller overlap cutoff  $(W_n=C_1N^{-1}n^{-7.5})$,
\item[--]  the nearest neighbour dissolution  $(W_n=C_2N^{-1}n^{-6})$ or exponential cutoff, and
\item[--] the red-wing Stark components dissolution  \\$(W_n=C_3N^{-2/3}n^{-4})$.
\end{itemize}
   Different partition functions and their dependence on  $N$ correspond
        to these cutoff methods.

     We have  used the last expression not only for H line series, but for
   smooth transition from Rydberg  line series to corresponding continua of all elements.
   Thus we obtain smooth transition from line series to the corresponding continua, but
  the state survival fractions are probably somewhat overestimated.
  It is extremely difficult to describe the real situation,
 because also a  complicated picture of Stark splitting of spectral lines must be taken into account.
 Although these problems are weakly connected with the LID computations, they are important for
 modelling of stellar atmospheres and spectra.

\section{Concluding remarks}
 Light-induced drift enables to explain and to model evolutionary generation of drastic isotopic anomalies in
quiescent CP stellar atmospheres, i.e. if stellar wind, meridional circulation and macroturbulence
are lacking. However, microturbulence, which
 is always present in the atmospheres, diminishes essentially
the equilibrium gradients of isotope ratios. Nevertheless, it has been succeeded to explain the main
regularity -- formation of essentially higher abundances of the heavier isotopes of heavy metals
on the example of mercury.

Further refinements of a number of physical process characteristics are needed.
The line profile functions and shifts,
 state survival fractions, impact cross-sections, Stark and hyperfine splitting of spectral lines
 and several other quantities should be improved
 to model evolutionary scenarios of isotope separation in
 CP stellar atmospheres.

\section*{Acknowledgements}
We are grateful to a referee for pointing on the necessity of discussion concerning
the possible influence of the MMM approach on our model computations.
We gratefully acknowledge the financial support by the Estonian Science Foundation grant  ETF7691.

 \bibliographystyle{elsarticle-harv}
\bibliography{sapar}

\end{document}